\documentclass[twocolumn,showpacs,preprintnumbers,amsmath,amssymb]{revtex4}
\usepackage{graphicx}
\usepackage{dcolumn}
\usepackage{bm}
\begin{document}
\title {Long wavelength behavior of the dynamical spin-resolved local-field
factor in a two-dimensional electron liquid}
\author {Zhixin Qian}
\affiliation{Department of Physics, and State Key Laboratory for
Mesoscopic Physics, Peking University, Beijing 100871, China}
\date{\today}
\begin{abstract}
The high frequency limits of the singular component
$A(\omega)$ of
the small wavevector expansion of the longitudinal (L) and transverse (T)
components of the spin-resolved
exchange-correlation kernel tensor 
$f_{xc, \sigma \sigma'}^{L,T} (q, \omega) 
= - v(q) G_{\sigma \sigma'}^{L,T} (q, \omega)$ in 
a two-dimensional isotropic electron liquid 
with arbitrary spin polarization are studied. Here 
$G_{\sigma \sigma'}^{L,T} (q, \omega)$ is the spin-resolved
local field factor, $v(q)$ is the Coulomb interaction in momentum space, and 
$\sigma$ denotes spin.
Particularly, the real part
of $A(\omega)$ is found to be logarithmically divergent at large $\omega$.
The large wavevector structure of the corresponding spin-resolved
static structure factor is also established.
\end{abstract}
\maketitle

\section{Introduction}

The spin-resolved local field factor $G_{\sigma \sigma'} (q, \omega)$
has often been used to include exchange-correlation (xc) effects
beyond the random phase approximation 
in many-electron systems \cite{singwi,ichimaru}.
It is also one of the key concepts in the
time-dependent (spin) density functional 
theory (TD(S)DFT) \cite{HK,KS,RG,GDP} .
In TD(S)DFT, however,
one frequently uses the equivalent xc kernel, defined as
\begin{eqnarray}  \label{f-G1}
f_{xc, \sigma \sigma'} (q, \omega) 
= - v(q) G_{\sigma \sigma'} (q, \omega) ,
\end{eqnarray}
instead. Here $v(q)$ is the Fourier transform of the Coulomb interaction.

Recently it has been found that $f_{xc, \sigma \sigma'} (q, \omega)$ has
quite surprising singular structure at small wavevector $q$. The singular
structure leads to invalidity of a conventional extension of  
local density approximation of the ground 
state density functional theory \cite{HK,KS}
to the TDSDFT. This problem of ultranonlocality in TDSDFT
has been recently solved in a scheme of time dependent spin current density 
functional theory (TDSCDFT) \cite{QCV}. In TDSCDFT, 
in addition to $f_{xc, \sigma \sigma'} (q, \omega)$ which coincides with the 
longitudinal component $f_{xc, \sigma \sigma'}^{L} ({\bf q}, \omega)$ 
of the xc kernel tensor $f_{xc, \sigma \sigma'}^{ij}
({\bf q}, \omega)$, the knowledge of the transverse
component $f_{xc,\sigma \sigma'}^T ({\bf q}, \omega)$ 
is also required. 
Note that we are only concerned with an isotropic
electron liquid and the directions are relative to the wavevector ${\bf q}$.
Thus, in general,
\begin{eqnarray}
f_{xc, \sigma \sigma'}^{L,T} (q, \omega) 
= - v(q) G_{\sigma \sigma'}^{L,T} (q, \omega) .
\end{eqnarray} 
Both $f_{xc, \sigma \sigma'}^L
(q, \omega)$ and $f_{xc, \sigma \sigma'}^T (q, \omega)$ have a singular 
structure at long wavelength \cite{QCV,QV},
\begin{eqnarray}
f_{xc, \sigma \sigma'}^{L,T} (q, \omega) = \frac{A(\omega)}{q^2}
\frac{\sigma \sigma' n^2}{4 n_{\sigma} n_{\sigma'}}
+ B_{\sigma \sigma'}^{L,T} (\omega) + O(q^2) ,
\end{eqnarray}
where $n_\sigma$ is the spin-resolved density
($\sigma =1$ for $\uparrow$-spin and $\sigma = -1$
for $\downarrow$-spin), and $n= n_\uparrow + n_\downarrow$
is the total density.
It has been shown $A(\omega)$ to be well 
behaved in three dimensions (3-D) \cite{QV}.
In this note, we point out that the real
part of $A(\omega)$, however, is divergent at high frequency 
in two dimensions (2-D). In other
words the third moment of the spin-density spin-density 
response function \cite{QV,goodman,liu} is infinity in 2-D. 
We further show that the imaginary part of $A(\omega)$ goes as a
constant at high frequency, which is consistent with the divergence
of $Re A(\omega)$, in accordance with Kramer-Kr\"onig (KK) relation,
\begin{eqnarray}  \label{KK}
Re A(\omega) = Re A(\infty) + P \int_{-\infty}^{\infty} 
\frac{d \omega'}{\pi} \frac{Im A (\omega')}{\omega' - \omega} .
\end{eqnarray}

\section{ Large wavevector limit of the 
spin-resolved static structure factor $S_{\sigma \sigma '} (k)$}

The third moment of the spin density spin density 
response function will be shown to be intrinsically related to
the spin-resolved static structure factor
$S_{\sigma \sigma'} (k)$. We start with an 
investigation of the large $k$ 
behavior of $S_{\sigma \sigma'} (k)$. 
$S_{\sigma \sigma'} (k)$ can be obtained
from the spin-resolved pair-correlation 
function $g_{\sigma \sigma'} (r)$ via the following well known
relation, 
\begin{eqnarray} \label{S-g}
S_{\sigma \sigma'} (k) - \frac{n_\sigma}{n} \delta_{\sigma \sigma'}
= \frac{n_\sigma n_{\sigma '}}{n} 
\int [ g_{\sigma \sigma'} (r) -1 ] e^{- i {\bf k} \cdot {\bf r}} d {\bf r} .
\end{eqnarray}
At large $k$, Eq. (\ref{S-g}) yields
\begin{eqnarray} \label{S}
\lim_{k \to \infty} k^3 S_{\sigma {\bar \sigma}} (k) 
= - \frac{4 \pi n_\uparrow n_\downarrow}{n a_0} 
g_{\sigma {\bar \sigma}} (0) ,
\end{eqnarray}
and
\begin{eqnarray} \label{S1}
\lim_{k \to \infty} k^5 [S_{\sigma \sigma} (k) - \frac{n_\sigma}{n} ]
= \frac{6 \pi n_\sigma^2}{n a_0} 
\frac{\partial^2}{\partial r^2} g_{\sigma \sigma}(r) |_{r =0} .
\end{eqnarray}
where ${\bar \sigma} = - \sigma$. Eq. (\ref{S}) was also established before
by Rajagopal and Kimball in Ref. \cite{R-K}, while Eq. (\ref{S1}) is new.  
In obtaining the above results, 
we have used the following properties of the short range 
structures of  
$g_{\sigma \sigma'} (r)$ \cite{kimball,rajagopal,carlsson},
\begin{eqnarray}
\frac{\partial}{\partial r} g_{\sigma {\bar \sigma}} (r) |_{r=0}
= \frac{2}{a_0} g_{\sigma {\bar \sigma}} (0) ,
\end{eqnarray}
and
\begin{eqnarray}
\frac{\partial^3}{\partial r^3} g _{\sigma \sigma} (r) |_{r=0} =
\frac{2}{a_0} \frac{\partial^2}{\partial r^2} 
g_{\sigma \sigma} (r) |_{r=0} ,
\end{eqnarray}
where $a_0$ is the Bohr radius. 

Recently an approximate scheme for 
the spin symmetric and antisymmetric local
field factors in 2-D was 
developed by Atwal et al in Ref. \cite{atwal1},
based on the exact limiting results and sum rules. One of the sum rules
employed in the scheme is the third moment sum rule. However, the fact
of infiniteness of the 
third moment, as will be demonstrated in the 
next section, was overlooked in Refs. \cite{atwal1,atwal2}.
To be more precise, the quantity $\lambda_a^{\infty} (r_s)$ 
in the above mentioned scheme,
which is defined in Eq. (26) in Ref. \cite{atwal1} (or equivalently
$I_2({\bf q} \to 0)$ defined in Eq. (25) in Ref. \cite{atwal2}), is infinite. 
In fact, Eq. (26) in Ref. \cite{atwal1} 
can be rewritten as
\begin{eqnarray} \label{lambdaa}
\lambda_a^{\infty} (r_s) = -8 \int_0^{\infty} 
d k k^2 S_{\uparrow \downarrow} (k) .
\end{eqnarray}
From Eq. (\ref{S}), one knows that $S_{\uparrow \downarrow} (k) \sim 1/k^3$ at
large $k$. Therefore the rhs of Eq. (\ref{lambdaa}) is divergent. The
infiniteness of $\lambda_a^{\infty} (r_s)$ seems 
not sufficiently realized in Refs. \cite{atwal1,atwal2} 
and it was evaluated numerically 
by use of the pair-correlation density $g(r)$ which is obtained at various
$r_s$ from QMC studies.

\section{High frequency limits of $A(\omega)$}

In this section, we will explicitly show the infiniteness of 
the third moment of the spin-density spin-density
response function, or equivalently the first moment of the spin-current
spin-current response function $M_{\sigma \sigma'}^{L,T} ({\bf q})$ \cite{QV}.
We start from a relation between the latter 
and $Re f_{xc, \sigma \sigma'}^{L,T} (q, \omega \to \infty)$,
\begin{eqnarray}   \label{sumrule}
v^\alpha (q) &+& Re f_{xc, \sigma \sigma'}^\alpha 
(q, \omega \to \infty)  \nonumber \\
&=& \frac{m^2}{n_\sigma n_{\sigma'}}   
\frac{M_{\sigma \sigma'}^\alpha ({\bf q})
- M_{\sigma \sigma'}^{\alpha(0)} ({\bf q})}{q^2} ,
\end{eqnarray}
where $\alpha = L, T$, and $v^L (q) = 2 \pi e^2 /q $, and $v^T (q) =0$.
$M_{\sigma \sigma'}^{\alpha(0)} ({\bf q})$ is the noninteracting
version of $M_{\sigma \sigma'}^\alpha ({\bf q})$. As in the case of
3-D \cite{QV}, $M_{\sigma \sigma'}^\alpha ({\bf q})$ has two parts
of contributions, one from kinetic energy, and the other from
electron interaction, 
\begin{eqnarray}
M^{\alpha}_{\sigma \sigma'} ({\bf q})
= [M^\alpha_{\sigma \sigma'} ({\bf q})]_{kin}
+ [M^\alpha_{\sigma \sigma'} ({\bf q})]_{int} .
\end{eqnarray}
Similar derivations to those in the case of the 3-D lead to
\begin{eqnarray}
[M^\alpha_{\sigma \sigma'} ({\bf q})]_{int}
&=& \frac{1}{m^2} q_\alpha^2 v(q) n_\sigma n_{\sigma'}   \nonumber \\
&+& \frac{1}{m^2} \biggl [ \delta_{\sigma \sigma'}
\sum_\tau \Gamma^\alpha_{\sigma \tau} (0) 
- \Gamma_{\sigma \sigma'}^\alpha ({\bf q}) \biggl ] ,
\end{eqnarray}
and
\begin{eqnarray}  \label{Mkin}
[M_{\sigma \sigma'}^\alpha ({\bf q}) ]_{kin}
= \frac{1}{S m^2} \delta_{\sigma \sigma'} 
(2 q_\alpha^2 +q^2 ) \langle T_\sigma \rangle,
\end{eqnarray}
where
\begin{eqnarray}
\Gamma_{\sigma \sigma'}^\alpha ({\bf q})
= - \frac{1}{S^2} \sum_{{\bf k} \neq {\bf q}}
k_\alpha^2 v( k) \langle \rho_\sigma ({\bf q} - {\bf k})
\rho_{\sigma'} ({\bf k} - {\bf q}) \rangle ,
\end{eqnarray}
and $\langle T_\sigma \rangle$ is the spin-resolved
kinetic energy, and $S$ is the area of the system.
At small wavevector, it can be shown that
\begin{eqnarray}  \label{Mint}
[M_{\sigma \sigma'}^\alpha ({\bf q}) ]_{int}
= && \frac{1}{m^2} n_\sigma n_{\sigma'} v(q) q_\alpha^2  \nonumber  \\
&& - \frac{n}{2 m^2 S} \sum_{{\bf k} \neq 0} 
\sigma \sigma' k^2 v(k) S_{\uparrow \downarrow} (k)  \nonumber \\
&&  + \beta^{L,T} \frac{ n_\sigma n_{\sigma'} q^2}{2 m^2}   \nonumber \\
&& \times \int d {\bf r} v({\bf r})  [1 - g_{\sigma \sigma'} (r) ] ,
\end{eqnarray}
where $\beta^L = -5/8$ and $\beta^T = 1/8$.
Substituting Eqs. (\ref{Mkin}) and (\ref{Mint}) into
Eq. (\ref{sumrule}), one has,
\begin{eqnarray} \label{A}
Re A(\omega \to \infty)= - \frac{2}{nS} \sum_{\bf k} k^2 
v(k) S_{\uparrow \downarrow} (k) ,
\end{eqnarray}
and
\begin{eqnarray}  \label{B}
B_{\sigma \sigma'}^{L,T} (\infty) = && \delta_{\sigma \sigma'}  
\alpha^{L,T} \frac{t_{c \sigma}}{n_\sigma}    \nonumber \\
&& + \frac{\beta^{L,T}}{2} \int d {\bf r} 
v (r) [ 1 - g_{\sigma \sigma'} (r) ] .
\end{eqnarray}
where $t_{c \sigma}$ is spin-resolved correlation 
kinetic energy for per particle,
and $\alpha^L = 3$, $\alpha^T =1$. 
Evidently $Re A (\omega \to \infty)$ approaches to infinity due to
the fact of  $S_{\uparrow \downarrow} (k) \sim 1/k^3 $ at large $k$
and subsequently the logarithmic divergence 
of the summation over ${\bf k}$ on the rhs
of Eq. (\ref{A}) . 

Next we turn to investigating the high frequency 
limit of the imaginary part of $A(\omega)$. Similar
calculations to those in the case of 3-D in 
Ref. \cite{QV} lead to the following exact identity,
\begin{eqnarray}  \label{ImA1}
Im A(\omega) = && \frac{2}{S^3 n^2}
\sum_{{\bf k}, {\bf k}'}  v({\bf k}) v ({\bf k}')
{\bf k} \cdot {\bf k}'    \nonumber \\
&& Im \langle \langle \rho_{\uparrow} (-{\bf k})
\rho_{\downarrow} ({\bf k}); \rho_{\uparrow} ({\bf k}')
\rho_{\downarrow} (- {\bf k}' ) \rangle \rangle ,
\end{eqnarray}
where $\rho_\sigma ({\bf k})$ is the spin-resolved density operator.
At large $\omega$, the four-point response function
coincides with that of a noninteracting electron gas,
which is given by
\begin{eqnarray}   \label{fourpoint} 
Im &&\langle \langle \rho_{\uparrow} (- {\bf k}) \rho_{\downarrow} ({\bf k});
 \rho_{\uparrow} ({\bf k}') 
\rho_{\downarrow} (-{\bf k}') \rangle \rangle ^{(0)}    \nonumber  \\
&& \stackrel{\omega \to \infty}{\to} - \pi \delta (\omega - k^2/m)
\delta_{{\bf k}, {\bf k}'} N_{\uparrow} N_{\downarrow} .
\end{eqnarray}
Substituting Eq. (\ref{fourpoint}) into Eq. (\ref{ImA1}), one has
\begin{eqnarray}
Im A (\omega) \stackrel{\omega \to \infty}{\to}
- \frac{2 \pi n_\uparrow n_\downarrow}{S n^2}
\sum_{\bf k} \delta (\omega - k^2/m) v^2 (k) k^2 ,
\end{eqnarray}
which can be further simplified as
\begin{eqnarray}   \label{ImA2}
Im A (\omega) \stackrel{\omega \to \infty}{\to}
- \frac{2 \pi^2 n_\uparrow n_\downarrow m e^4}{n^2} .
\end{eqnarray} 
Therefore, surprisingly, though reasonably in accordance
with the divergent structure of $Re A (\omega \to \infty)$,
$Im A(\omega) $ goes as a constant at large $\omega$. 
In fact, with KK relation of Eq. (\ref{KK}),
the limiting structure $Im A (\omega)$ in Eq. (\ref{ImA2}) yields
\begin{eqnarray}
Re A(\omega) \stackrel{\omega \to \infty}{\to}
\frac{4 \pi n_\uparrow n_\downarrow m e^4}{n^2} \ln \omega .
\end{eqnarray}
In obtaining the above result, we have used the fact of
$Re A(0) =0 $, which is proved for 3-D 
case in Ref. \cite{QCV,QV}, and evidently 
also holds for 2-D case.

\section{Conclusions}

Several exact results for the spin-resolved 
static structure factor and the dynamical local field
factor of an arbitrary spin-polarized electron liquid 
in two dimensions have been established
in this paper. Particularly, we have found that, 
at small wavevector, the corresponding third moment
of the spin-density spin-density response function
is infinite, and the exchange-correlation
kernel is divergent logarithmically at high frequency. 

Note added in proof: After the paper was finished, it was
found that the divergence of
the quantity $\lambda_a^{\infty}(r_s)$ defined in Ref. \cite{atwal1,atwal2}
was also noticed by Dr. P. Gori-Giorgi.

We thank Drs. G. Vignale, P. Gori-Giorgi, 
A. K. Rajagopal, and M. Polini for stimulating discussions.

\end{document}